\documentclass{article}
\usepackage{spconf,amsmath,graphicx}

\usepackage{bm,bbm}
\usepackage{amsfonts,amssymb,mathtools,fourier,times}
\usepackage{subcaption}
\usepackage{hyperref}
\usepackage{version}
\excludeversion{hide}

\newtheorem{Theorem}{Theorem}

\newcommand{\e}{{\mathrm e}}
\newcommand{\im}{{\mathrm i}}
\newcommand{\dd}[1]{\, {\mathrm d}{#1}}
\newcommand{\bdm}{\begin{displaymath}}
\newcommand{\edm}{\end{displaymath}}
\newcommand{\cond}{\,|\,}
\newcommand{\define}{\, := \,}
\newcommand{\F}{{\mathcal{F}}}

\newcommand{\inner}[2]{\left\langle {#1}, {#2} \right\rangle}
\newcommand{\M}{{\mathcal{M}}}
\newcommand{\pv}{\,\, \mathrm{p.v.}\!\!}
\newcommand{\Riesz}{{\mathcal{R}}}
\newcommand{\sgn}{\operatorname{sgn}}
\newcommand{\transp}{{\scriptscriptstyle{\mathsf{T}}}}
\newcommand{\bb}{{\bm{b}}}
\newcommand{\bs}{{\bm{s}}}

\newcommand{\bu}{{\bm{u}}}
\newcommand{\bx}{{\bm{x}}}

\newcommand{\bxi}{{\bm{\xi}}}
\newcommand{\blambda}{{\bm{\lambda}}}

\newcommand{\bi}{{\mathbbm{i}}}
\newcommand{\bj}{{\mathbbm{j}}}
\newcommand{\bk}{{\mathbbm{k}}}

\newcommand{\Z}{{\mathbb{Z}}}
\newcommand{\Rf}{{\mathbb{R}}}
\newcommand{\CC}{C\nolinebreak\hspace{-.04em}\raisebox{.3ex}{\scriptsize\bf +}\nolinebreak\hspace{-.10em}\raisebox{.3ex}{\scriptsize\bf +}\,}

\title{Monogenic Wavelet Scattering Network for Texture Image Classification}
\name{{Wai Ho Chak} and Naoki Saito \thanks{$^{\star}$
    This research was partially supported by the US National Science Foundation
    grants DMS-1912747, CCF-1934568; the US Office of Naval Research
    grant N00014-20-1-2381.}}
\address{University of California, Davis\\
  Department of Mathematics\\
  One Shields Avenue, Davis, CA 95616 USA}
\begin{document}
%
\maketitle
\begin{abstract}
  The scattering transform network (STN), which has a similar structure as that
  of a popular convolutional neural network except its use of predefined 
  convolution filters and a small number of layers, can generates a robust
  representation of an input signal relative to small deformations.
  We propose a novel \emph{Monogenic Wavelet Scattering Network} (MWSN)
  for 2D texture image classification through a cascade of monogenic wavelet
  filtering with nonlinear modulus and averaging operators by replacing the
  2D Morlet wavelet filtering in the standard STN.
  Our MWSN can extract useful hierarchical and directional features with
  interpretable coefficients, which can be further compressed by PCA and
  fed into a classifier. Using the CUReT texture image database, we demonstrate
  the superior performance of our MWSN over the standard STN.
  This performance improvement can be explained by the natural extension of 1D
  analyticity to 2D monogenicity.
\end{abstract}
\section{Introduction}
\label{sec:intro}
The \emph{scattering transform network} (STN) has an architecture similar to
a popular convolutional neural network (CNN).
The latter, in particular, its ``deep'' version called Deep Neural Network (DNN),
is known for its ability to extract hierarchical and critical features for many
applications such as image classification and facial recognition when large
training samples are available~\cite{lecun2015deep, sun2013deep}.
Despite its popularity, the CNN/DNN tends to overfit its model for a dataset of
small size.
The STN, on the other hand, works well without a gigantic number of examples,
and requires no optimization
(e.g., stochastic gradients) to learn convolution filters from data since it
uses predefined wavelet filters. 
In addition, the STN typically operates with a small number of layers, say,
two or three, which is quite a contrast to the DNN.
Yet, Mallat showed that the STN representation of an input signal provides
a translation invariant representation when the scale tends to infinite
resolution, and it is Lipschitz continuous under non-uniform
translation~\cite{mallat2012group}. Furthermore, Bruna and Mallat
demonstrated the power of the STN using image classification examples~\cite{bruna2011classification, bruna2013invariant}.

A typical software implementation of the STN, e.g.,
the Kymatio package~\cite{andreux2020kymatio}, uses the Morlet wavelet filter
as its base wavelet filter.
Although the Morlet wavelets are only approximately
\emph{analytic}~\cite{LILLY-OLHEDE-2}, they have been used in the
\emph{analytic wavelet transform} (AWT).
The AWT provides interpretable multiscale instantaneous magnitude and phase
information, which is crucial for 1D signal analysis. 
When one wants to analyze 2D input images, the concept of analyticity needs to
be properly extended: simply considering the tensor product of the 1D AWT would
not be sufficient.
A properly extended concept of 1D analytic signal to higher dimension
is the so-called \emph{monogenic signal} proposed by Felsberg and Sommer~\cite{felsberg2001monogenic}; see also \cite{MonogenicSignalTheory} for the comprehensive
review.
Based on the monogenic signal theory, Olhede and Metikas proposed the
\emph{monogenic wavelet transform} (MWT)~\cite{olhede2009monogenic}
generalizing the 1D AWT.
The MWT inherits those desirable properties of AWT for higher-dimensional
signals, i.e., provides their multiscale instantaneous magnitude, phase, and
orientation information, which would lead to better
interpretability of the important features of high-dimensional input signals.
Moreover, the isotropic MWT magnitude is invariant to rotations,
and its orientation is covariant to rotations~\cite{olhede2009monogenic}. 

Hence, it is quite natural and in fact inviting to adopt the MWT as the base
wavelet transform in the STN framework instead of the 2D Morlet wavelet
transform when handling 2D images. By doing so, we can capture 2D geometric
features more efficiently than what the Morlet wavelets-based STN could provide.

\section{Monogenic Signals and MWT}
\label{sec:monogenic_sec}
In this section, after briefly reviewing the concept of an
analytic signal and the AWT, we will describe the definition and properties of a
monogenic signal and the MWT, which are necessary to discuss our \emph{Monogenic
Wavelet Scattering Network} (MWSN) in Section~\ref{sec:MWSN}.

A signal is said to be \emph{analytic} if it is a complex-valued function that
has no negative frequency components~\cite{GABOR, VILLE}.
The analytic representation provides useful information such as
instantaneous amplitude and phase. Hence, if a real-valued signal,
say, $g(x) \in L^2(\Rf)$ is given,
it is of interest to generate its analytic representation $g^+(x)$
whose real component is $g(x)$.
It is well known that its imaginary component can be obtained by
the \emph{Hilbert transform} of $g(x)$~\cite{GABOR, VILLE, HahnHilbert1996},
i.e.,
\bdm
g^+(x) \define g(x) + \im g^{(1)}(x), \quad
g^{(1)}(x) \define \frac{1}{\pi} \pv \int_{-\infty}^\infty \dfrac{g(y)}{x - y} \dd{y},
\edm
where $\pv$\, means the Cauchy principal value.

The notion of wavelet is also necessary. A \emph{wavelet} (or also known as
\emph{mother wavelet}) $\psi(x) \in L^2(\Rf)$ is a function whose translated
and dilated versions allow one to perform time-scale/frequency analysis of
a signal of interest~\cite[Chap.~4]{MALLAT-BOOK3}.
It has to satisfy the \emph{admissibility condition:}
\bdm
\int_{\Rf} \dfrac{| \Psi(\xi) |^2}{|\xi|} \dd{\xi} < \infty, \quad
\int_\Rf |\psi(x)|^2 \dd{x} = 1 ,
\edm
where $\Psi(\xi)$ is the Fourier transform of $\psi(x)$.
In 1D, the \emph{continuous wavelet transform} (CWT) of a signal $g \in L^2(\Rf)$ with respect to the mother wavelet $\psi$ is given by
\bdm
W_\psi g(a, b) \define \dfrac{1}{\sqrt{a}} \int_\Rf g(t) \overline{\psi\left(\dfrac{t - b}{a} \right)} \dd{t} ,
\edm
for any $a \in \Rf_+ \define \{ x \in \Rf \cond x > 0 \}$, $b \in \Rf$.
The CWT defines the AWT when $\psi$ is analytic.

We can extend the concept of analyticity to higher dimensions.
With the fact that a 1D analytic signal is a boundary value (at the real
axis) of a complex-valued analytic function that satisfies the
\emph{Cauchy-Riemann equations} in the upper half of the complex plane~\cite{HahnHilbert1996, MonogenicSignalTheory}, we can find the analogue of the 2D analytic
signal. 
A popular choice for generalizing the Cauchy-Riemann equations to higher
dimensions is \emph{the Riesz system} (of PDEs)~\cite{STEIN-WEISS} in the upper
half-space $\Rf^3_+ \define \{ \bx \in \Rf^3 \cond x_3 > 0 \}$.
Any solution of the Riesz system in $\Rf^3_+$ is said to be
a \emph{monogenic}~\cite{felsberg2001monogenic}, and its boundary value at
$x_3=0$ is called \emph{monogenic signal}.

Next we need to introduce the \emph{quaternion}, which has the form $e \define e_0 + e_1 \bi + e_2 \bj + e_3 \bk \in \mathbb{H}$, where $\mathbb{H}$ is the 4D real associative algebra of the quaternions, $\bi, \bj, \bk$ are the quaternion units, and $e_l \in \Rf$, $l = 0, \cdots, 3$. The quaternion units satisfy the following multiplication rules: $\bi^2 = \bj^2 = \bk^2 = -1$; $\mathbbm{ij} = -\mathbbm{ji} = \bk$; $\mathbbm{jk} = -\mathbbm{kj} = \bi$; $\mathbbm{ki} = -\mathbbm{ik} = \bj$. 
In addition, if $e \in \mathbb{H}$, then we have $\overline{e} = e_0 - e_1 \bi - e_2\bj - e_3 \bk$, and
$\| e \| = \sqrt{e \overline{e}} = \sqrt{e_0^2 + e_1^2 + e_2^2 + e_3^2}$.
The notion of quaternion is needed when we introduce the definition of the
Fourier transform of a 2D signal by identifying the quaternion unit $\bj$ as
the imaginary unit $\im$ as well as the Riesz transform. The Fourier transform
of a 2D signal $g(\bx) \in L^2(\Rf^2)$ is given by
\bdm
\F g (\bxi) = G(\bxi) \define \int_{\Rf^2} g(\bx) \e^{-2 \pi \bj \bxi^\transp \bx} \dd{\bx}.
\edm
Let us define the \emph{Riesz transform} \cite{STEIN-WEISS, HahnHilbert1996}
of $g(\bx) \in L^2(\Rf^2)$ by
\bdm
\Riesz g(\bx) \define \bi  \Riesz_1 g(\bx) + \bj  \Riesz_2 g(\bx),
\, \Riesz_l g(\bx) \define (r_l \ast g)(\bx), \, l = 1, 2,
\edm
where $r_l(\bx) \define x_l/(2\pi \| \bx \|^3)$ is the \emph{Riesz kernel}.
Note that the Fourier transform of the Riesz kernel
$\F r_l(\bxi) = -\bj  \xi_l / \| \bxi \|$
can be contrasted with that of the 1D Hilbert transform kernel:
$\F h(\xi) = - \im \sgn(\xi) = -\im \xi / | \xi |$ where
$h(x) \define 1/(\pi x)$.
The monogenic signal $g^+$ of the signal $g$ 
is now defined by introducing an operator 
$\M^+ \define \mathrm{I} + \Riesz$ \cite{felsberg2001monogenic, MonogenicSignalTheory}
such that 
\bdm
\M^+ g(\bx) = g^+(\bx) \define g(\bx) + \Riesz g(\bx) .
\edm
A monogenic signal can be decomposed into ``instantaneous'' amplitude, phase,
and orientation components~\cite{MonogenicSignalTheory} as follows:
\begin{equation}
  g^{+}(\bx) = A(\bx) \left(\cos \phi(\bx) + \nu(\bx) \sin \phi(\bx)\right) ,
\label{eqn:mono-decomp}
\end{equation}
where $A(\bx) \define \| g^{+}(\bx) \|$,
$\phi(\bx) \define \cos^{-1} ( g(\bx)/\| g^{+}(\bx) \| )$, and
$\nu(\bx) \define \Riesz g(\bx) / \| \Riesz g(\bx) \|$
describe the \emph{amplitude}, \emph{phase}, and
\emph{phase direction (or orientation)} information locally at $\bx \in \Rf^2$,
respectively.
Note that $\| \Riesz g(\bx) \| \define \sqrt{|g^{(1)}(\bx)|^2 + |g^{(2)}(\bx)|^2}$
whereas $\| g^{+}(\bx) \| \define \sqrt{|g(\bx)|^2 + \|\Riesz g(\bx)\|^2}$.
Finally, the MWT is defined as the 2D CWT with a monogenic mother wavelet.

\section{Monogenic Wavelet Scattering Network}
\label{sec:ST-MWSN}
\subsection{Basics of scattering transform network}
Let $Q_m$ be a discrete finite rotation group in $\Rf^2$ at layer $m$.
Note that in any STN architecture, $m$ is typically set at $2$, i.e.,
it is quite shallow compared to DNNs.
Denote $\Lambda_m \define Q_m \times \Z$ the $m$-th layer index set consisting
of the rotation $q \in Q_m$ and the scale $j \in \Z$.
Let $\lambda_m = (q, j)  \in \Lambda_m $ be the index for a multiscale
directional wavelet filter at layer $m$. We can obtain such wavelet filters
by dilating and rotating a mother wavelet $\psi$.
The generator or the multiscale directional wavelet corresponds to the index
$\lambda_m = (q,j)$ is
\bdm
\psi_{\lambda_m}(\bx) \define 2^{2j} \psi(2^j q^{-1} \bx) .
\edm
Note that we assume the generator $\psi_{\lambda_m} \in L^1(\Rf^2) \cap L^2(\Rf^2)$.
It is known as a frame atom and corresponds to a receptive field of a CNN
\cite{WIATOWSKI-BOLCSKEI-DEEP-THEORY-IT}.

To be more precise, let $f \in L^2(\Rf^2)$, and let us define
a \emph{translation} operator $T_{\bb}f(\bx) \define f(\bx - \bb)$ and
an \emph{involution} operator operator $If(\bx) \define \overline{f(-\bx)}$.
Then, a \emph{frame atom} is defined by
$\psi_{\bb, \lambda_m} \define T_\bb I \psi_{\lambda_m}$. Note that
$\inner{f}{\psi_{\bb, \lambda_m}} = f \ast \psi_{\lambda_m}(\bb)$.
A \emph{contraction} operator $M_m$ which is \emph{Lipschitz continuous}
can also be defined, and satisfies $M_m f(\bx) = 0 \Rightarrow f(\bx) = 0$.
A particular choice of $M_m$ is the \emph{modulus} operator, i.e.,
$M_m f(\bx) \define | f(\bx) |$.
We define an operator $U_m: \Lambda_m \times L^2(\Rf^2) \rightarrow L^2(\Rf^2)$
from layer $m-1$ to layer $m$ such that
\begin{equation}
U_m[\lambda_m] f(\bx) \define M_m (f \ast \psi_{\lambda_m})(r_m \bx) .
\label{eqn:u1-def}
\end{equation}
$r_m \geq 1$ represents a \emph{subsampling rate}.
Thus we have a \emph{path} of indices
$\blambda \in \Lambda_m \times \cdots \times \Lambda_1$ such that
\begin{equation}
U[\blambda] f(\bx) \define U_m[\lambda_m] U_{m-1}[\lambda_{m-1}] \cdots U_1[\lambda_1] f (\bx) .
\label{eqn:u2-def}
\end{equation}
For each layer $m$, we define the operators $S_m$ and $\Phi_m$ to generate
the output feature vectors (or coefficients) of the MWSN for a given
input signal $f(\bx)$:
\begin{eqnarray}
  \label{eqn:sm-def}
  S_m[\blambda]f(\bx) &\define& (\varphi_m \ast U[\blambda] f) (r'_m \bx), \\ 
  \Phi_m f(\bx) &\define& \left\{ S_m[\blambda]f(\bx) \right\}_{\blambda \in  \Lambda_m \times \cdots \times \Lambda_1}, \nonumber
\end{eqnarray}
where $\varphi_m$ is an averaging function, e.g., the father wavelet of a
certain scale corresponding to the mother wavelet $\psi$ and
$r'_m \geq 1$ provides yet another subsampling opportunity after this averaging
process. Note that for $m=0$, we set $S_0[\emptyset]f(\bx) = S_0f(\bx) \define (\varphi_0 \ast f)(r'_0 \bx)$.

\subsection{Monogenic wavelet scattering network}
\label{sec:MWSN}
\begin{figure}
\centering
\includegraphics[width=0.49\textwidth]{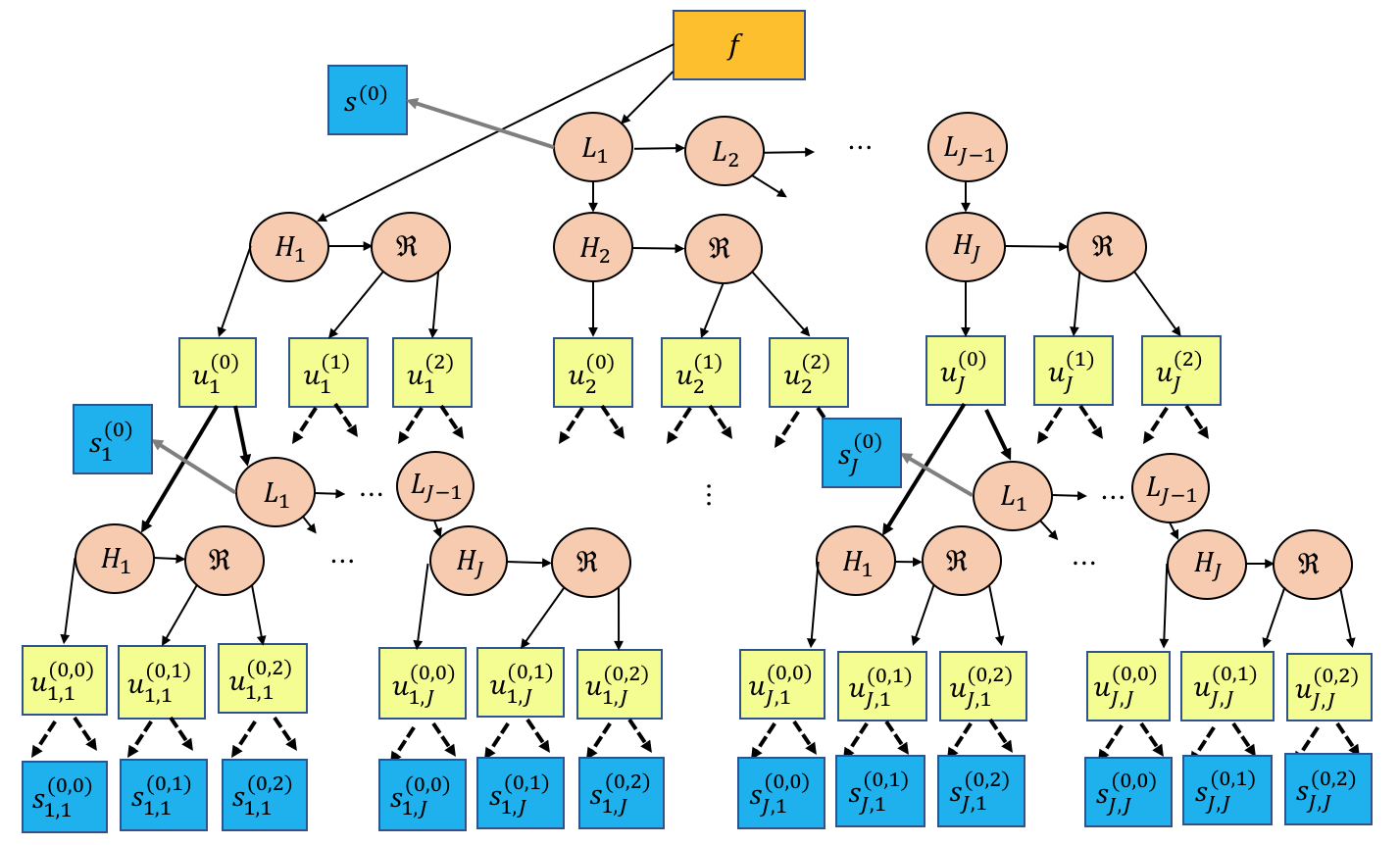}
\caption{The MWSN architecture}
\label{fig:MWSN}
\end{figure}

The visual description of the Monogenic Wavelet Scattering Network (MWSN)
is shown in Fig.~\ref{fig:MWSN}. 
For the MWT implementation, we follow the strategy of Soulard and
Carr\'e~\cite{SOULARD-CARRE-2018} as follows.
First, let us define a Gaussian high-pass filter $H$ in the 2D Fourier domain:
\bdm
H(\bxi) \define 1 - \e^{-\frac{{\| \bxi \|}^2}{2}} .
\edm
Then the Gaussian high-pass filter at scale $j$, denoted by $H_j$,
is defined by
\begin{equation}
  H_j(\bxi) \define H(2^{j-1} \bxi), \quad j \in \{1, \ldots, J\},
\label{eqn:Hj}
\end{equation}
where we typically use $J=4$ in practice.
A low-pass filter $L_j$ at scale $j$ can be defined as
\bdm
L_j(\bxi) \define \sqrt{1- (H_j(\bxi))^2} .
\edm
In the MWSN framework, the MWT filter bank has intensive interaction with the
feature from the previous layer.
In Fig.~\ref{fig:MWSN}, the salmon-pink disks represent operators,
i.e., $H_j$, $L_j$, and $\Riesz$.
Note that the convolution with the father wavelet $\varphi_m$ at the $m$th layer
in Eq.~\eqref{eqn:sm-def} in the conventional STN corresponds to the low-pass
filtering with $L_1$ at every layer in the MWSN. 
The zeroth ($m=0$) layer output is indicated by the blue box $S_0f(\bx)$
($\bs^{(0)}$ for short) after the low-pass filtering with $L_1$ of the input image
$f$ followed by subsampling. The superscript $(0)$ indicates the
isotropic filtering is applied. 
In the first ($m=1$) layer, the vectors $U_1[\lambda_1]f(\bx)$,
$\lambda_1 \in \Lambda_1$ in the conventional STN of Eq.~\eqref{eqn:u1-def}
is now denoted by the vectors in the yellow boxes, $\bu^{(l)}_j$,
$j \in \{1, \ldots, J\}$, $l \in \{0, 1, 2\}$, where $j$ is the scale parameter,
and $l \in \{0, 1, 2\}$ indicates the isotropic component, the vertical and
horizontal Riesz components obtained by $\Riesz_1$, $\Riesz_2$, respectively. 
The output vectors of the first layer, indicated by blue boxes such as
$\bs^{(0)}_1$ and $\bs^{(0)}_J$, are obtained by subsampling $\bu^{(l)}_j$, 
low-pass filtering with $L_1$, and yet another subsampling.
Note that the other first-layer outputs, i.e.,
$\bs^{(0)}_j$,$j \in \{2, \dots, J-1\}$
and $\bs^{(l)}_j$, $l \in \{1, 2\}$, $j \in \{1, \ldots, J\}$ are omitted due to
the crowded graphics.
Now, in the second ($m=2$) layer, the vector $U[\blambda]f(\bx) = U[\lambda_2]U[\lambda_1]f(\bx)$ 
in the conventional STN of Eq.~\eqref{eqn:u2-def} is denoted by
$\bu^{(l_1,l_2)}_{j_1,j_2}$, $j_k \in \{1, \ldots, J\}$, $l_k \in \{0, 1, 2\}$,
$k=1,2$,
where $l_1, j_1$ indicate the inherited first layer path information whereas
$l_2, j_2$ are the parameters specified in the second layer.
The outputs of the second layer are again obtained by applying the same procedure
as the first layer to $\bu^{(l_1,l_2)}_{j_1,j_2}$, which are indicated by blue boxes
$\bs^{(l_1,l_2)}_{j_1,j_2}$. 
Finally, the arrows in this diagram show the flow of the data; in addition,
the thick arrows indicate that the subsampling operations are performed before
reaching the destination disks or boxes while their color (gray or black)
suggests that a potentially different subsampling rate can be set.

\section{Numerical Results and Discussion}
\label{sec:result}
Using the CUReT texture image dataset~\cite{dana1999reflectance},
we evaluated the classification performance of our MWSN and compared with
that of the standard 2D STN based on Morlet wavelets~\cite{bruna2011classification, bruna2013invariant}, i.e., the Kymatio package implemented by Andreux
et al.~\cite{andreux2020kymatio} in the Python programming language.
All the other codes we used for our experiments are based on the Julia
programming language~\cite{JULIA}.
For each of 61 texture classes in the CUReT dataset, we selected 92 texture
images that were cropped to retain a 200 $\times$ 200 central region and
converted to grayscale. Hence, the total number of available texture images is
$5,612$.
We set the maximum scale parameter $J=4$ in Eq.~\eqref{eqn:Hj} in the MWSN,
which exactly corresponds to $J=3$ in the Kymatio package. For both methods,
we only used the second layer outputs because they contain the most relevant
information. The subsampling rates in the MWSN were set to $2$ regardless of
the layers while we used the default values in the Kymatio-STN.
Since the Kymatio-STN allows the users to choose the number of orientations
of the Morlet wavelets, we tried the number of orientations
$L=2, 4, 6, 8$. In each case, we used the PCA implemented in the
\texttt{MultivariateStats.jl} package~\cite{MultivariateStats} to reduce the
dimension of the coefficient/feature vectors of the MWSN and the Kymatio-STN.
After some experiments, we decided to use the top $30$ PCA coordinates for
all cases.
Then, those coordinates were fed to the Support Vector Machine (SVM) classifier
(of a polynomial kernel of degree 1) implemented in the \texttt{LIBSVM.jl}
package~\cite{LIBSVM} that is based on the \CC library~\texttt{LIBSVM}~\cite{SVM}.
Table~\ref{table: acc} lists the average test accuracy by repeating two-fold
cross validation 10 times for each case along with the coefficient/feature vector
dimension before the PCA was applied.
\begin{table}
\begin{center}
\begin{tabular}{| c | r | c | c |}
\hline
        & Dimension  & Classification  \\
Method  & before PCA & Accuracy        \\
\hline\hline
               MWSN & 90,000  &{\bf 97.34\%} \\
  \hline
Kymatio-STN (L = 2) & 11,875  & 94.50\% \\
  \hline
Kymatio-STN (L = 4) & 38,125  & 96.18\% \\
  \hline
Kymatio-STN (L = 6) & 79,375  & 96.56\% \\
  \hline
Kymatio-STN (L = 8) & 135,625 & 96.61\% \\
  \hline
\end{tabular}
\end{center}
 \caption{Table of average test accuracy over 10 experiments} 
 \label{table: acc}
\end{table}
The best accuracy achieved by our proposed MWSN is due to the natural extension
of analyticity in 1D to monogenicity in 2D. The CWT with Morlet wavelets
retain less properties than the MWT due to the ``leak'' of
the energy to the negative frequency range. Together with the fact that the
Riesz kernels are effective 2D edge detectors, fewer contextual directions in
the MWSN can still capture sufficient textural information to achieve the better
classification result than the Kymatio-STN does at least for this particular
CUReT texture image dataset.

\begin{figure}
\begin{subfigure}{0.15\textwidth}
  \centering\includegraphics[width=\textwidth]{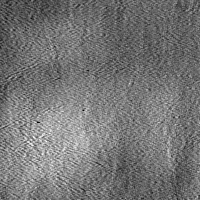}
  \caption{An input texture image}
  \label{fig:orig_img}
\end{subfigure}
\begin{subfigure}{0.35\textwidth}
  \centering\includegraphics[width=\textwidth]{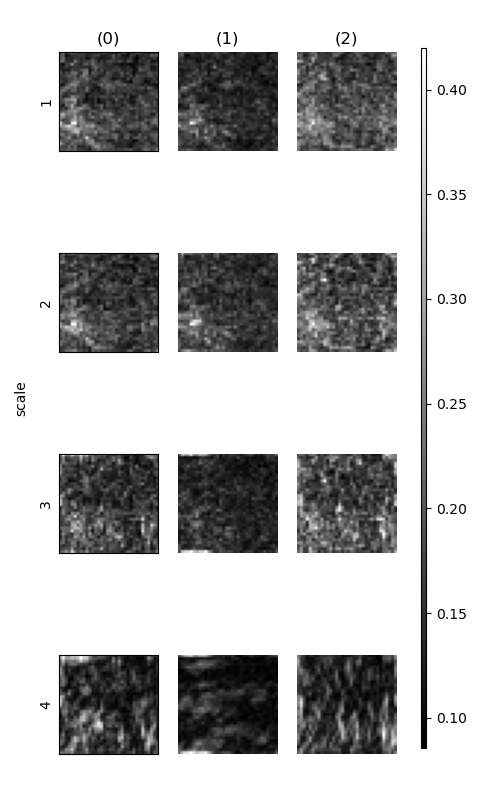}
  \caption{The first layer output}
  \label{fig:curet_layer1}
\end{subfigure}
\caption{A CUReT image and its MWSN first layer output}
\end{figure}

\begin{figure}
\centering
\includegraphics[width = 0.5\textwidth]{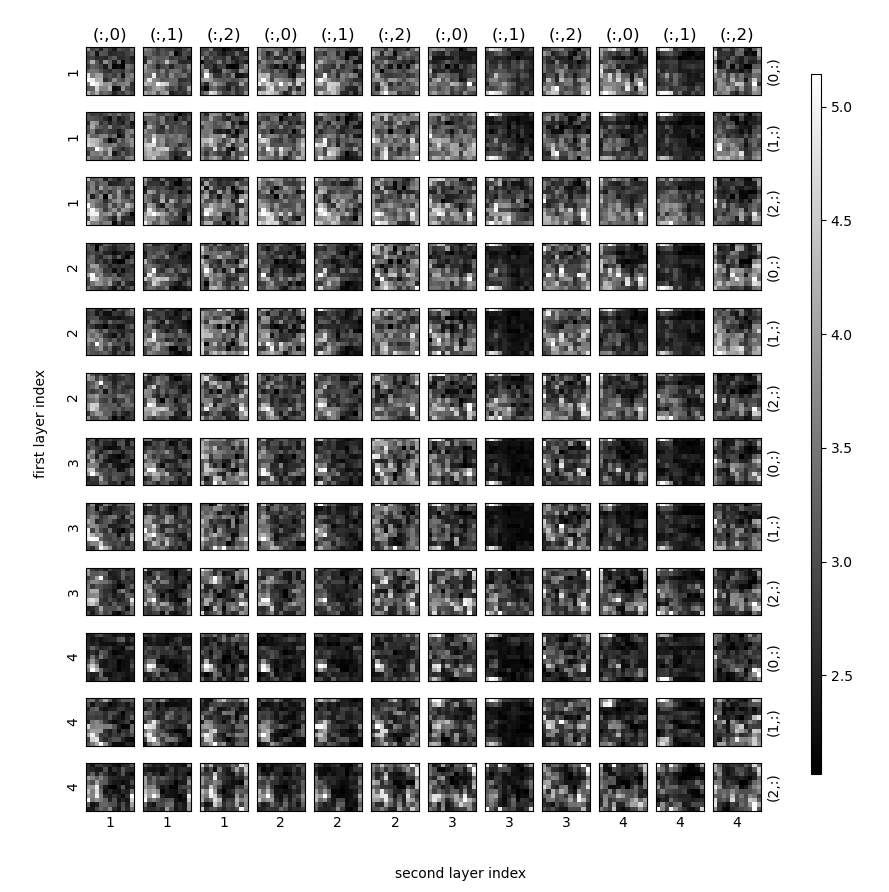}
\caption{The MWSN second layer output of the image shown in Fig.~\ref{fig:orig_img}}
\label{fig:curet_layer2}
\end{figure}

The output coefficients of our proposed MWSN are also ``interpretable.''
Interpretability was rarely considered in the earlier studies on texture
classification. Fig.~\ref{fig:curet_layer1} and Fig.~\ref{fig:curet_layer2}
clearly illustrate the orientation information of an image captured by the Riesz
transforms, $\Riesz_1$ and $\Riesz_2$, respectively.
In Fig.~\ref{fig:curet_layer2}, the $(i,j)$th block contains $25 \times 25$ MWSN
coefficients $\bs^{(l_1, l_2)}_{j_1, j_2}$ with $l_1 = (i-1) \mod 3$,
$l_2 = (j-1) \mod 3$, $j_1 = \lfloor (i-1)/3 \rfloor + 1$, and $j_2=\lfloor (j-1)/3 \rfloor + 1$, where $i, j \in \{1, \ldots, 12\}$.
As we traverse from left to right at each row of Fig.~\ref{fig:curet_layer2},
we see more intricate texture information is captured.

\begin{hide}
Furthermore, we can also visualize which MWSN coefficients are important
for correct classification if we replace SVM by the logistic regression
classifier~\cite[Chap.~3]{HASTIE-TIB-WAINWRIGHT} implemented in
\texttt{GLMNet.jl}~\cite{GLMNet}. This is because we can identify
the important PCA components indicated by the significant components of the
regression coefficient vectors, and then project them back to the original
MWSN coordinates.
\end{hide}

To improve the interpretability of the MWSN coefficients that are important
for classification, we have three plans: 1) replace SVM by the logistic
regression classifier~\cite[Chap.~3]{HASTIE-TIB-WAINWRIGHT} so that we
can pinpoint such MWSN coefficients; 2) convert the second layer coefficients
$\bs^{(l_1,l_2)}_{j_1,j_2}$ into the instantaneous amplitude, phase, and orientation
representation via Eq.~\eqref{eqn:mono-decomp} before applying the PCA;
and 3) replace the PCA by the Local Discriminant Basis (LDB)
method~\cite{SAITO-COIF-JMIV, SAITO-COIF-GESHWIND-WARNER} since the latter can
directly extract features that are helpful for classification instead of
extracting high variance features by the PCA.

\newpage

\bibliographystyle{IEEEbib}
\bibliography{reference}
\end{document}